\documentclass[%
 reprint,
superscriptaddress,
 amsmath,amssymb,
 aps,
]{revtex4-2}

\usepackage{graphicx}
\usepackage{dcolumn}
\usepackage{bm}
\usepackage{amsmath}
\usepackage{hyperref}
\usepackage{color}
\usepackage{multirow}
\usepackage{array}
\usepackage{booktabs}
\usepackage[T1]{fontenc}
\usepackage{mathptmx}
\usepackage{upgreek}

\begin{document}

\preprint{APS/123-QED}

\title{Sympathetic cooling of a large ${}^{113}\mathrm{{Cd}}^{+}$ ion crystal with ${}^{40}\mathrm{{Ca}}^{+}$ in a linear Paul trap}
 
\author{S. N. Miao}
 \email{These authors contributed equally.}
\affiliation{
 State Key Laboratory of Precision Measurement Technology and Instruments, Key Laboratory of Photon Measurement and Control Technology of Ministry of Education, Department of Precision Instruments, Tsinghua University, Beijing 100084, China
}

\author{H. R. Qin}
 \email{These authors contributed equally.}
\affiliation{
 State Key Laboratory of Precision Measurement Technology and Instruments, Key Laboratory of Photon Measurement and Control Technology of Ministry of Education, Department of Precision Instruments, Tsinghua University, Beijing 100084, China
}
\affiliation{
 Department of Physics, Tsinghua University, Beijing 100084, China
}

\author{N. C. Xin}
\affiliation{
 State Key Laboratory of Precision Measurement Technology and Instruments, Key Laboratory of Photon Measurement and Control Technology of Ministry of Education, Department of Precision Instruments, Tsinghua University, Beijing 100084, China
}

\author{Y. T. Chen}
\affiliation{
 State Key Laboratory of Precision Measurement Technology and Instruments, Key Laboratory of Photon Measurement and Control Technology of Ministry of Education, Department of Precision Instruments, Tsinghua University, Beijing 100084, China
}

\author{J. W. Zhang}
 \email{zhangjw@tsinghua.edu.cn}
\affiliation{
 State Key Laboratory of Precision Measurement Technology and Instruments, Key Laboratory of Photon Measurement and Control Technology of Ministry of Education, Department of Precision Instruments, Tsinghua University, Beijing 100084, China
}

\author{L. J. Wang}
 \email{lwan@tsinghua.edu.cn}
\affiliation{
 State Key Laboratory of Precision Measurement Technology and Instruments, Key Laboratory of Photon Measurement and Control Technology of Ministry of Education, Department of Precision Instruments, Tsinghua University, Beijing 100084, China
}
\affiliation{
 Department of Physics, Tsinghua University, Beijing 100084, China
}

\date{\today}

\begin{abstract}
We have sympathetically cooled and crystallized ${}^{113}\mathrm{{Cd}}^{+}$ ions with 
laser-cooled ${}^{40}\mathrm{{Ca}}^{+}$ ions, and directly observed the complete large bicrystal structure in a linear Paul trap. The large two-component crystal contains up to $3.5\times10^3$ ${}^{40}\mathrm{{Ca}}^{+}$ ions and $6.8\times10^3$ ${}^{113}\mathrm{{Cd}}^{+}$ ions. The temperature of the crystallized ${}^{113}\mathrm{{Cd}}^{+}$ ions was measured to be as low as 41 mK with a large mass ratio. 
Moreover, we have studied several properties and structures of the ultracold sample. The factors affecting the sympathetic cooling effect were studied, including the electrical parameters and the number ratio between laser-cooled ions and sympathetically-cooled ions. The results of this paper enrich the experimental research of large two-component ion crystals, and the ultracold sample of ${}^{113}\mathrm{{Cd}}^{+}$ ions makes it possible to further improve the accuracy of the cadmium-ion microwave frequency standard.
\end{abstract}

\maketitle


\section{\label{sec:level1}Introduction}
Since laser cooling was first proposed in the late 1970s \cite{h1975cool,a1978trap} and laser-cooled sodium atoms were first cooled to sub-millikelvin level in the 1980s \cite{chu1986ex},
the technology of laser-cooling has received extensive attention and applications in Bose-Einstein condensation, precision measurement, and searching for new physics \cite{wie1999atom, cron2009opt, nova2018sea}. As a result, a new field of ultralow temperature physics was born. 
However, many atoms (molecules, ions) cannot be directly cooled by laser because they do not have suitable electric dipole transitions or suitable wavelength lasers are not available. For example, laser cooling of anions is still a challenge today \cite{tang2019}. Another limitation of laser cooling is that the ultracold atoms ensemble will quickly heat up as soon as the cooling laser is turned off, which limits the quantum control and precise measurement of the low-temperature ensemble.

In the past few decades, to overcome the limitations of laser cooling, the technology of sympathetic cooling, using one species of low-temperature particle as the coolant to cool down another species of particle, has been introduced into the experiment. For example, buffer-gas cooling is one of the most common forms of sympathetic cooling. However, when talking about sympathetic cooling, it usually means using laser-cooled ultracold particles as coolant to sympathetically cool another species of particles through collisions and other interactions. Sympathetic cooling between neutral atoms makes it possible to research more on degenerate Fermi gases and molecular Bose–Einstein condensates \cite{demarco2002spin,truscott2001observation,geist1999sympathetic,myatt1997production}. Besides, hybrid traps were invented to sympathetically cool ions with ultracold atoms \cite{smith2005cold}.
Since ions can be confined in ion traps spatially and can be cooled down with just one laser beam, ions are often used as the coolants to cool another species of ions. Sympathetic cooling of trapped ions in the Penning trap and Paul trap were first realized in 1980 \cite{d1980} and 1992 \cite{b1992m}.
So far, the technology of sympathetic cooling can be used to cool any atomic or molecular ion (singly charged) with the mass between 1 and 470 amu \cite{krems2009cold}, even charged proteins (cytochrome c molecules, charge $+17e$) as heavy as 12390 amu can be sympathetically cooled to below 0.75 K \cite{off2008t}. 

To date, the sympathetic cooling of few-ion systems has been extensively studied \cite{wubbena2012sympathetic} and widely used. For example, the $\mathrm{{}^{27}Al^+}$ optical atomic clock sympathetically cooled by $\mathrm{{}^{25}Mg^+}$ has reached an extremely high-stability below $10^{-18}$ \cite{brewer2019al+}. The qubit of $\mathrm{{}^{171}Yb^+}$ sympathetically cooled by $\mathrm{{}^{138}Ba^+}$ has the longest coherence-time up to around 5500 s \cite{wang2021single}. 
As for large-number  ($\textgreater$ 1000) ion systems, they are often applied in the fields of precision measurement and spectroscopy because of their high signal-to-noise ratio. For example, over 5000 trapped $\mathrm{{}^{9}Be^+}$ ions sympathetically cooled by $\mathrm{{}^{26}Mg^+}$ have been used for a 303-MHz frequency standard with high accuracy \cite{bollinger1991303}. The sympathetic cooling of molecular ions with a large number of ions is also usually used to study chemical reactions \cite{molhave2000formation,h2000phd,roth2006production}, photofragmentation \cite{offenberg2009measurement}, and molecular spectra \cite{koelemeij2007vibrational,bertelsen2006rotational,koelemeij2007blackbody}. 

Since 2019, our group has been committed to applying the sympathetic cooling technology to the cadmium-ion microwave frequency standard in order to reduce the limitation of Dick effect and the uncertainty in the second-order Doppler frequency shift (SODFS). In 2019, we used ${}^{24}\mathrm{{Mg}}^{+}$ as the coolant to cool the large ${}^{113}\mathrm{{Cd}}^{+}$ cloud \cite{zuo2019direct}. However, the reaction between $\mathrm{{Mg}}^{+}$ and background $\mathrm{{H}}_{2}$ turns the $\mathrm{{Mg}}^{+}$ into dark ions ($\mathrm{{MgH}}^{+}$), which limits the sympathetic cooling efficiency. In 2021, we replaced ${}^{24}\mathrm{{Mg}}^{+}$ with ${}^{40}\mathrm{{Ca}}^{+}$ as the coolant.
Although the mass ratio of 0.354 between ${}^{40}\mathrm{{Ca}}^{+}$ and ${}^{113}\mathrm{{Cd}}^{+}$ limits the  sympathetic cooling efficiency, sympathetic cooling of ${}^{113}\mathrm{{Cd}}^{+}$ ions was still preliminarily achieved \cite{han2021toward}. However, the large two-component ion crystals in the Paul trap are still lacking in thorough research. The research on the properties and structures of the ultracold sample and the optimization of parameters still need to be carried out in the experiment. Considering the limitations of the experiment, molecular dynamics (MD) simulation is introduced to compare with experiments and optimize experimental parameters.

In this paper, we report on the production of an ultracold large bicrystal containing ${}^{113}\mathrm{{Cd}}^{+}$ and ${}^{40}\mathrm{{Ca}}^{+}$ ions. Trapping of these ions is straightforward using a linear Paul trap. The temperature of the  sympathetically-cooled ${}^{113}\mathrm{{Cd}}^{+}$ crystal was measured to be 41 mK with as many as 6800 ions and a large mass ratio. The lifetime of the ${}^{113}\mathrm{{Cd}}^{+}$ crystal was estimated to be about $2\times10^4$ seconds, which is three times longer than that in the laser cooling scheme. Moreover, we studied several properties and structures of the ultracold sample containing single species and double species of ions. The factors affecting the sympathetic cooling effect were studied in detail, including the electrical parameters and the number ratio between laser-cooled (LC) ions and sympathetically-cooled (SC) ions.

\section{Experimental setup}
\begin{figure}[t!]
\centering
\includegraphics[width=7cm]{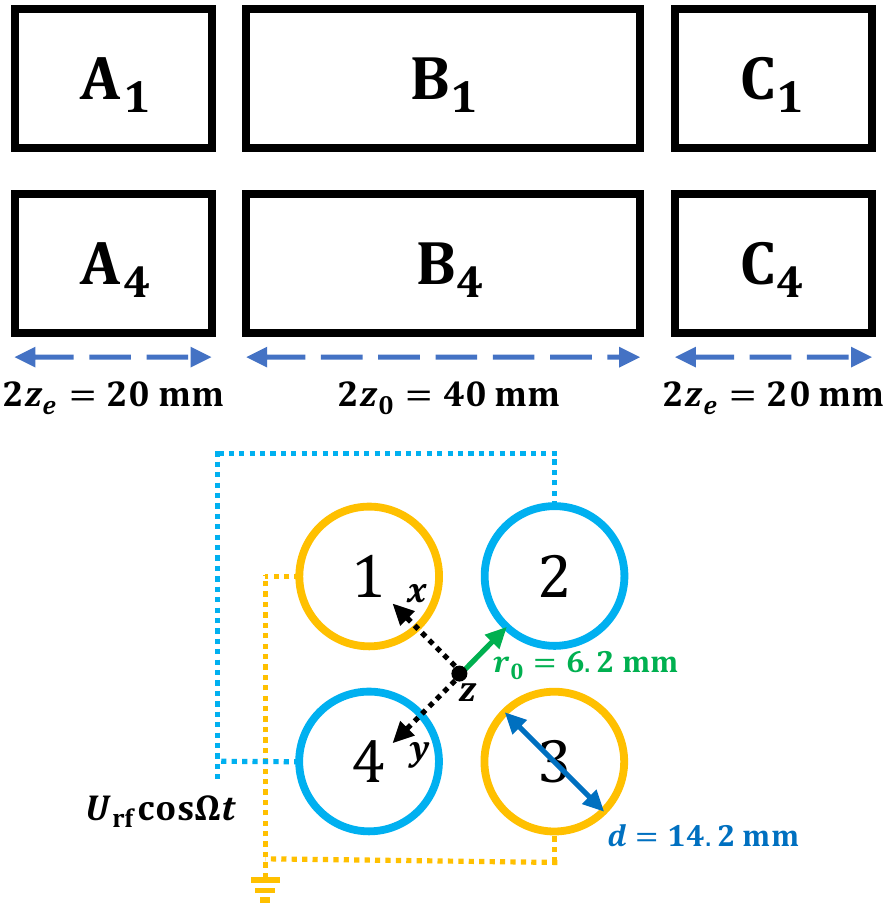}
\caption{\label{fig:fig_01} (Color online) Schematic of the linear quadrupole Paul trap. The rod diameter ($d$) and the inner radius ($r_0$) are 14.2 and 6.2 mm, respectively. The central trap region ($2z_0$) is 40 mm. The origin of the three-dimensional (3D) coordinate axis coincides with the geometric center of the ion trap.}
\end{figure}

The experimental setup has been described in more detail elsewhere \cite{miao2021precision,han2021toward}; here, a brief description of the linear Paul trap suffices (Fig. \ref{fig:fig_01}). It consists of four rod electrodes (1,2,3,4) with a diameter of $d=14.2$ mm, and each rod is segmented into three parts (A,B,C). The minimum distance between the nodal line of the trap and the electrode surfaces is $r_0=6.2$ mm. The lengths of the trapping part (B) and the remaining parts (A,C) of each rod are $2z_0=40$ mm and $2z_e=20$ mm, respectively. Confinement of ions is achieved by applying a radio frequency (RF) voltage $U_{\mathrm{rf}} \cos (\Omega t)$ to one pair of diagonal electrode rods ($\mathrm{B}_2$,$\mathrm{B}_4$) and a DC voltage $U_{\mathrm{end}}$ to end electrodes (A,C). 

\begin{figure}[t!]
\centering
\includegraphics[width=8cm]{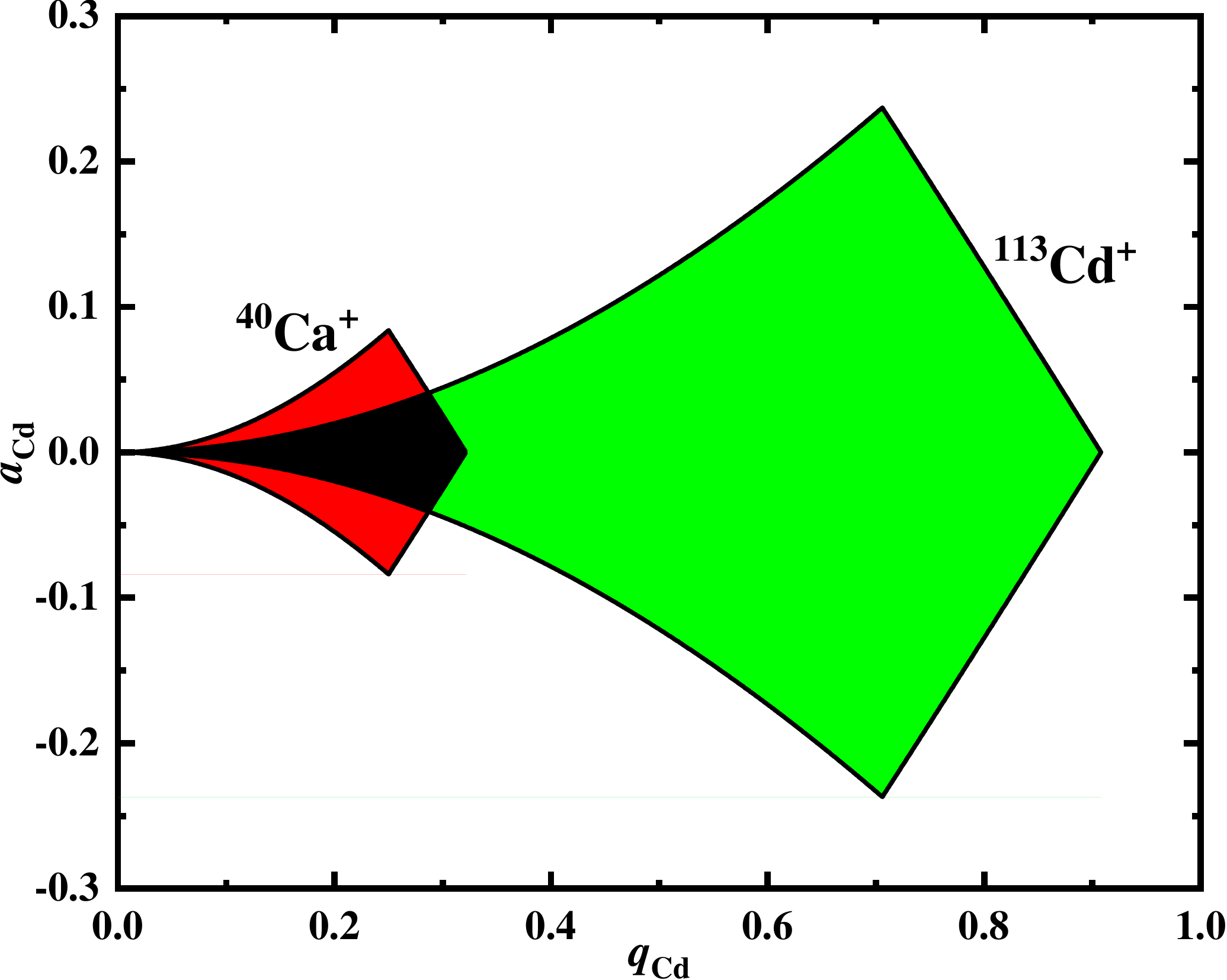}
\caption{\label{fig:fig_02} (Color online) The lowest stability region for the radial plane, where $a$ and $q$ are present in the charge-mass ratio of ${}^{113}\mathrm{{Cd}}^{+}$. Green part is the lowest stability region for ${}^{113}\mathrm{{Cd}}^{+}$; red part is the lowest stability region for ${}^{40}\mathrm{{Ca}}^{+}$; black part is the lowest stability region both for ${}^{113}\mathrm{{Cd}}^{+}$ and ${}^{40}\mathrm{{Ca}}^{+}$.}
\end{figure}

The radial motion of a single ion trapped in a quadrupole Paul trap can be described by the Mathieu equation \cite{paul1958ion}:
\begin{equation}
\begin{aligned}
\label{equ01}
\frac{\mathrm{d}^{2} \boldsymbol{r}}{\mathrm{d} \xi^{2}}+\left(a-2 q \cos 2 \xi\right) \boldsymbol{r}=0,
\end{aligned}
\end{equation}
where $\boldsymbol{r}=x, y$. $a$, $q$, and $\xi$ denote three dimensionless parameters given by
\begin{equation}
\begin{aligned}
\label{equ02}
a_{x}&=a_{y}=\frac{4 Q \kappa U_{\mathrm{end}}}{M \Omega^{2} z_{0}^{2}}, \\
\quad q_{x}&=-q_{y}=\frac{2 Q U_{\mathrm{rf}}}{M \Omega^{2} r_{0}^{2}}, \\
\quad \xi&=\frac{\Omega t}{2},
\end{aligned}
\end{equation}
where $M$ and $Q$ denote the mass and charge of the ion, $\kappa=0.12$ is the axial geometrical factor related to the configuration of the quadrupole trap. To confine ${}^{40}\mathrm{{Ca}}^{+}$ and ${}^{113}\mathrm{{Cd}}^{+}$ ions in the trap simultaneously, the parameters $a$ and $q$ should be taken into account carefully. The values of $a$ and $q$ differ for different ion species depending on their charge-mass ratio. For ${}^{113}\mathrm{{Cd}}^{+}$ and ${}^{40}\mathrm{{Ca}}^{+}$, $a_{\mathrm{Cd}} / a_{\mathrm{Ca}}=q_{\mathrm{Cd}} / q_{\mathrm{Ca}}=(M_{\mathrm{Cd}} / M_{\mathrm{Ca}})^{-1}=0.354$. Therefore, $U_{\mathrm{rf}}$ and $U_{\mathrm{end}}$ need be adjusted carefully considering the overlapping stability region (Fig. \ref{fig:fig_02}). In addition, the depth of the potential well in the center of the trap needs to be greater than 3 eV in order to trap ions stably, and the adiabatic condition ($q \textless 0.3$) should be met. 

The linear Paul trap with the frequency $\Omega/2\pi=1.961$ MHz is placed in an ultrahigh vacuum chamber ($\textless 10^{-9}$ Pa). In our experiment, the typical amplitudes of the RF voltage and DC voltage are $U_{\mathrm{rf}}=150$ V and $U_{\mathrm{end}}=10$ V, respectively, yielding values of $q_{\mathrm{Ca}}$ = 0.124, $q_{\mathrm{Cd}}$ = 0.044, $a_{\mathrm{Ca}}\approx0$, and $a_{\mathrm{Cd}}\approx0$.

\section{Two-component ion crystal}
In order to obtain two-component ion crystals in the low-temperature phase, we usually apply the following operations:

First, neutral $\mathrm{{Ca}}$ atoms are evaporated by heating the $\mathrm{{Ca}}$ oven for 27 s at 6 A and photo-ionized by two lasers at the wavelengths of 423 nm ($\mathrm{{Ca}}\ 4s^2\ {}^1S_0\rightarrow4s4p\ {}^1P_1$) and 374 nm ($\mathrm{{Ca}}\ 4s4p\ {}^1P_1\rightarrow \mathrm{Continuum}$).
Natural abundance of solid $\mathrm{{Ca}}$ is placed in the atomic oven, of which ${}^{40}\mathrm{{Ca}}$ accounts for up to 97\%.
In order to cool ${}^{40}\mathrm{{Ca}}^{+}$ ions, we use two lasers, $\lambda = 397$ nm for Doppler cooling (${}^{40}\mathrm{{Ca}}^+\ 4s\ {}^2S_{1/2}\rightarrow4p\ {}^2P_{1/2}$) and $\lambda = 866$ nm for repumping the ions from the dark state (${}^{40}\mathrm{{Ca}}^+\ 3d\ {}^2D_{3/2}\rightarrow4p\ {}^2P_{1/2}$), frequency stabilized by a multichannel wavelength meter (WSU-4060). The fluorescence signal from the ${}^{40}\mathrm{{Ca}}^{+}$ ions is simultaneously recorded with a photomultiplier tube (PMT, Hamamatsu H8259-01) and an electron-multiplying charge coupled device (EMCCD). By scanning the frequency of the cooling laser (397 nm) and reducing the amplitude of RF voltage, we first cool down the calcium ions and obtain the ultracold ${}^{40}\mathrm{{Ca}}^{+}$ crystal.

Subsequently, ${}^{113}\mathrm{{Cd}}^{+}$ ions are produced by ionizing neutral atoms evaporated from an isotopically enriched ${}^{113}\mathrm{{Cd}}$ oven with the 228-nm laser beam. During ${}^{113}\mathrm{{Cd}}^{+}$ loading, the 214-nm cooling laser (${}^{113}\mathrm{{Cd}}^+\ {}^2S_{1/2}\ F=1, m_F=+1\rightarrow{}^2P_{3/2}\ F=2, m_F=+2$) and the microwave radiation of 15.2 GHz (${}^{113}\mathrm{{Cd}}^+\ {}^2S_{1/2}\ F=0\rightarrow{}^2S_{1/2}\ F=1$) are applied simultaneously to cool down the ions. Meanwhile, the filter before the PMT is replaced with a suitable one so that the fluorescence signal of ${}^{113}\mathrm{{Cd}}^{+}$ ions can be observed. Then we repeatedly scan the frequency of the calcium-ion cooling laser (397 nm) and reduce RF voltage until there is a sharp mutation in the intensity of the fluorescence signal of ${}^{113}\mathrm{{Cd}}^{+}$ ions on the PMT, which means ${}^{113}\mathrm{{Cd}}^{+}$ ions are sympathetically cooled to the crystal phase by ${}^{40}\mathrm{{Ca}}^{+}$ ions. At this time, the 214-nm laser beam is only used as the detection light for ${}^{113}\mathrm{{Cd}}^{+}$ ions and does not affect the structure and temperature of the ${}^{113}\mathrm{{Cd}}^{+}$ crystal, even blocked by an optical shutter.
The frequency of the 397-nm cooling laser is held constant at a red detuning of 20 MHz from the ${}^{40}\mathrm{{Ca}}^{+}$ resonance. Under this condition the two-component ion crystal is stable in the ultracold crystal phase.

\begin{figure}[t!]
\centering
\includegraphics[width=8cm]{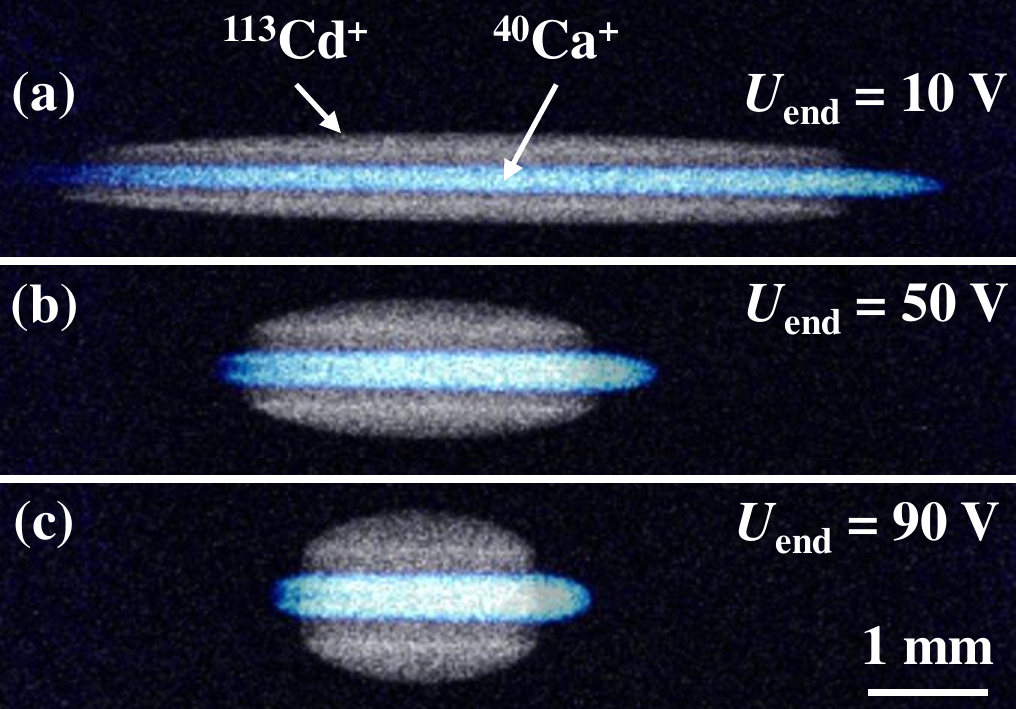}
\caption{\label{fig:fig_03} (Color online) EMCCD images of a large ${}^{40}\mathrm{{Ca}}^{+}$-${}^{113}\mathrm{{Cd}}^{+}$ bicrystal under the influence of three different DC voltages (a) $U_\mathrm{end}=10$ V (b) $U_\mathrm{end}=50$ V (c) $U_\mathrm{end}=90$ V. The crystal contains approximately $3.5(0.8)\times10^3$ ${}^{40}\mathrm{{Ca}}^{+}$ ions and $6.8(0.9)\times10^3$ ${}^{113}\mathrm{{Cd}}^{+}$ ions. The RF voltage is fixed to 150 V. The trap axis $z$ is horizontal. For ${}^{40}\mathrm{{Ca}}^{+}$, the cooling laser beam direction is to the right, while the direction is the opposite for the ${}^{113}\mathrm{{Cd}}^{+}$ cooling laser beam.}
\end{figure}

\begin{figure}[h!]
\centering
\includegraphics[width=8cm]{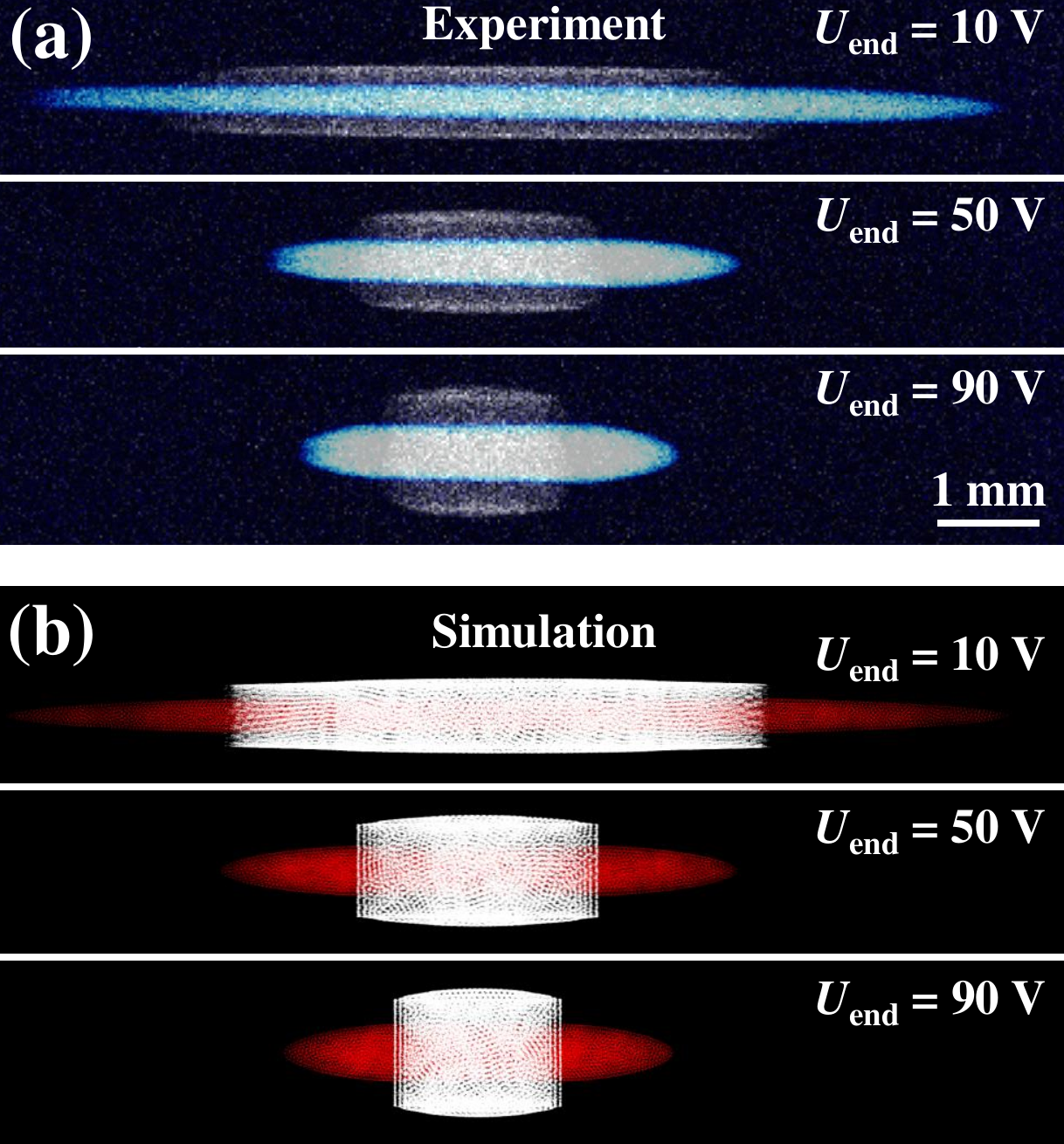}
\caption{\label{fig:fig_04} (Color online) (a) Experimental images of a large ${}^{40}\mathrm{{Ca}}^{+}$-${}^{113}\mathrm{{Cd}}^{+}$ bicrystal at three different DC voltages under the influence of a fixed RF voltage $U_\mathrm{rf}=195$ V. The number of ${}^{40}\mathrm{{Ca}}^{+}$ ions is estimated to be about $1.6(0.3)\times10^4$ with a spheroid fit to the envelope. The ${}^{113}\mathrm{{Cd}}^{+}$ ions form a thin shell outside of the ${}^{40}\mathrm{{Ca}}^{+}$ crystal instead of a hollow ellipsoid. (b) Simulated images of the two-component ion crystal at three different DC voltages under the influence of a fixed number of ${}^{40}\mathrm{{Ca}}^{+}$ ions, which reproduce the observed images in (a). The numbers of ${}^{40}\mathrm{{Ca}}^{+}$ (red) and ${}^{113}\mathrm{{Cd}}^{+}$ (white) ions are $1.6\times10^4$ and $4.6\times10^3$, respectively.}
\end{figure}

Through filters for different wavelengths in front of the EMCCD, images of ${}^{40}\mathrm{{Ca}}^{+}$ and ${}^{113}\mathrm{{Cd}}^{+}$ ions are obtained separately. Then we synthesize the images of the two types of ions, considering the chromatic aberration effect of the lens for light of different wavelengths. Figures \ref{fig:fig_03} and \ref{fig:fig_04}(a) show the experimental images of large two-component ion crystals containing ${}^{40}\mathrm{{Ca}}^{+}$ and ${}^{113}\mathrm{{Cd}}^{+}$ ions under the influence of different number ratios. In Fig. \ref{fig:fig_03}, we show the structures of the same group of ions under different DC voltages. The ${}^{40}\mathrm{{Ca}}^{+}$ crystal locates in the center of the trap with an ellipsoidal structure, whereas the ${}^{113}\mathrm{{Cd}}^{+}$ ions form a hollow ellipsoid surrounding ${}^{40}\mathrm{{Ca}}^{+}$. It is clear that increasing DC voltage decreases the axial size, but increases the radial size of the bicrystal. In order to determine the number of ions, the volume and density of the ion crystal are required. Considering that the ${}^{113}\mathrm{{Cd}}^{+}$ crystal is hollow, we obtain the separation ratio of the two species of ions by the zero-temperature charged-liquid model \cite{o1981pla,wine1987ion,h2000phd,h2001st}, which is
\begin{equation}
\label{equ0r}
\frac{r_{\mathrm{{Cd}}^{+}}}{r_{\mathrm{{Ca}}^{+}}}=\sqrt{\frac{M_{\mathrm{{Cd}}^{+}}}{M_{\mathrm{{Ca}}^{+}}}},
\end{equation}
where $r_{\mathrm{{Cd}}^{+}}$ denotes the radius of the inner surface of the ${}^{113}\mathrm{{Cd}}^{+}$ crystal, $r_{\mathrm{{Ca}}^{+}}$ the radius of the outer surface of the ${}^{40}\mathrm{{Ca}}^{+}$ crystal.
With Eq. (\ref{equ0r}) and the fitting dimensions of the ${}^{40}\mathrm{{Ca}}^{+}$ and ${}^{113}\mathrm{{Cd}}^{+}$ crystals, the volumes for ${}^{40}\mathrm{{Ca}}^{+}$ and ${}^{113}\mathrm{{Cd}}^{+}$ are estimated to be $V_{\mathrm{{Ca}}^{+}}=0.26(0.06)$ $\mathrm{mm}^{3}$ and $V_{\mathrm{{Cd}}^{+}}=1.44(0.18)$ $\mathrm{mm}^{3}$, respectively.

The density of ions of type $i$ is given by \cite{wine1987ion,h2001st}:
\begin{equation}
\label{equ0n}
n_i=\frac{\epsilon_0 U_{\mathrm{rf}}^2}{M_i \Omega^2 r_0^4},
\end{equation}
where $\epsilon_0$ denotes the permittivity of vacuum. In Fig. \ref{fig:fig_03}, the RF voltage is 150 V with a driving frequency of $\Omega=2\pi\times 1.961$ MHz.
The ion densities for ${}^{40}\mathrm{{Ca}}^{+}$ and ${}^{113}\mathrm{{Cd}}^{+}$ are calculated to be $n_{\mathrm{{Ca}}^{+}}=2.22\times10^{13}$ $\mathrm{m}^{-3}$ and $n_{\mathrm{{Cd}}^{+}}=4.71\times10^{12}$ $\mathrm{m}^{-3}$. Therefore, the numbers of ions in the two-component crystal are estimated to be approximately $N_{\mathrm{{Ca}}^{+}}=3.5(0.8)\times10^{3}$ and $N_{\mathrm{{Cd}}^{+}}=6.8(0.9)\times10^{3}$, respectively.

As the number ratio of $N_{\mathrm{{Ca}}^{+}}/N_{\mathrm{{Cd}}^{+}}$ increases, the ${}^{113}\mathrm{{Cd}}^{+}$ ions form a thin shell outside of the ${}^{40}\mathrm{{Ca}}^{+}$ crystal instead of a hollow ellipsoid, which is shown in Fig. \ref{fig:fig_04}(a).
The pictures of the large bicrystal in Fig. \ref{fig:fig_04}(a), containing ${}^{40}\mathrm{{Ca}}^{+}$ and ${}^{113}\mathrm{{Cd}}^{+}$ ions at a RF amplitude of $U_\mathrm{rf}=195$ V, are snapshots from a series obtained while varying the DC voltage. Changing $U_\mathrm{end}$ does not lead to variation in ion density according to Eq. (\ref{equ0n}).
The envelope of the ${}^{40}\mathrm{{Ca}}^{+}$ crystal can be fitted with an ellipsoid, and the volume is estimated to be approximately $V_{\mathrm{{Ca}}^{+}}=0.72(0.14)$ $\mathrm{mm}^3$. The ion density is $n_{\mathrm{{Ca}}^{+}}=2.26\times10^{13}$ $\mathrm{m}^{-3}$, yielding the ion number of $N_{\mathrm{{Ca}}^{+}}=1.6(0.3)\times10^4$. However, the determination of the volume of the ${}^{113}\mathrm{{Cd}}^{+}$ crystal has become a lot more complicated because of its structure. In this situation, the number of SC ${}^{113}\mathrm{{Cd}}^{+}$ ions is estimated by performing MD simulations and varying the number of ${}^{113}\mathrm{{Cd}}^{+}$ ions until the observed EMCCD images are reproduced. Details of our MD simulation model are given in \cite{xin2021research,miao2021research}; here, a brief description suffices. The MD model is based on Newton’s classical equations of motion. The ions confined in the trap are mainly affected by the trapping force of the potential field, the Coulomb interaction force, the random collision force, and the laser force. The equations of motion are solved using the second-order, stable, time-reversible leapfrog algorithm with an adaptive step size. During the simulation, the number of ${}^{40}\mathrm{{Ca}}^{+}$ ions is fixed to 16000. The graphs of Fig. \ref{fig:fig_04}(b) show simulated images of sympathetic crystallized 4600 ${}^{113}\mathrm{{Cd}}^{+}$ ions by laser-cooled 16000 ${}^{40}\mathrm{{Ca}}^{+}$ ions produced by different DC voltages, which reproduce the observed EMCCD images in Fig. \ref{fig:fig_04}(a) well. Therefore, the number of ${}^{113}\mathrm{{Cd}}^{+}$ ions is estimated to be approximately $4.6(0.2)\times10^3$.

\begin{figure}[t!]
\centering
\includegraphics[width=8cm]{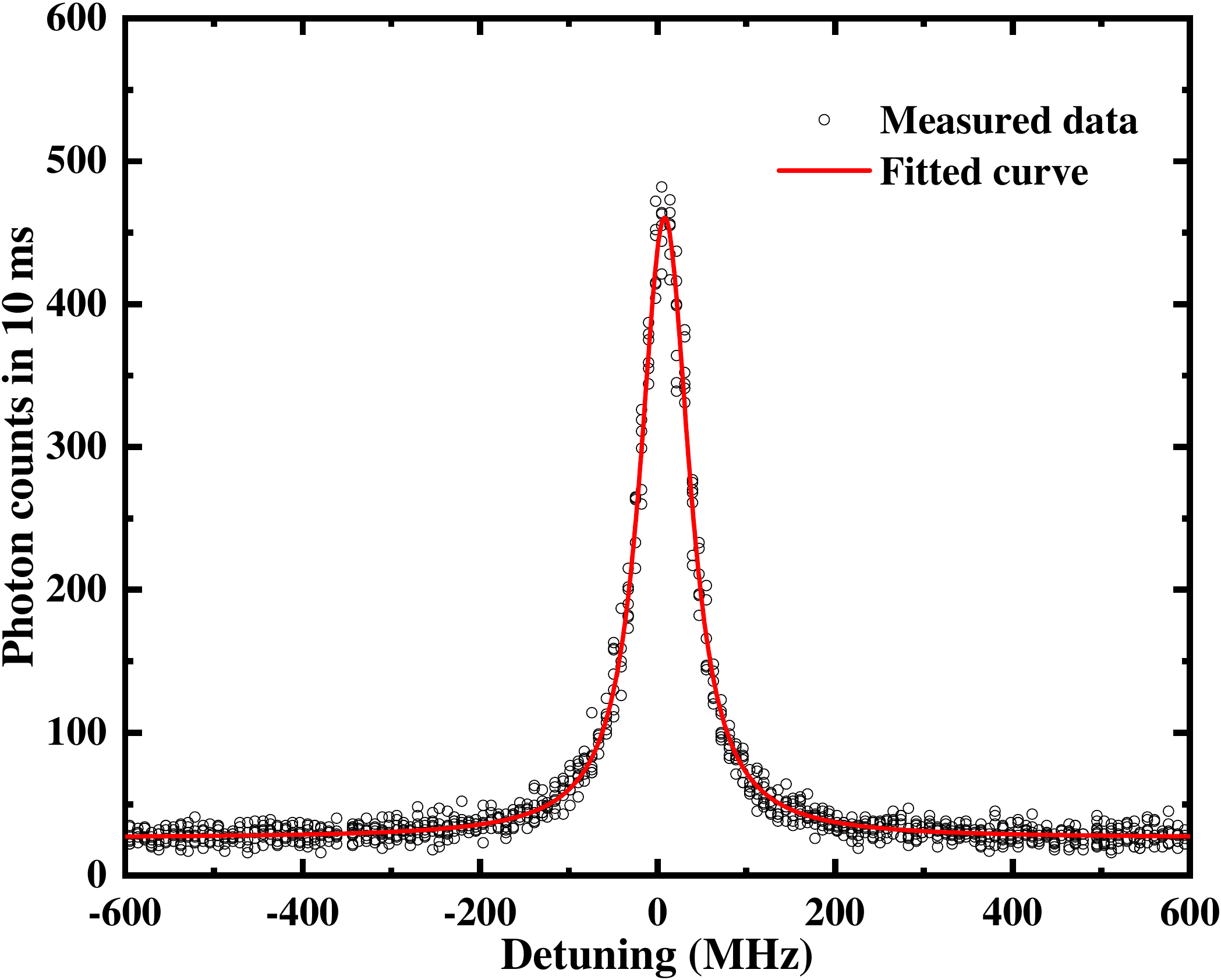}
\caption{\label{fig:fig_05} (Color online) Typical temperature measurement result of SC ${}^{113}\mathrm{{Cd}}^{+}$ ions. The line profile is fitted with a Voigt function under the influence of a fixed Lorentz linewidth of 60.13 MHz, which is the nature linewidth of the $\mathrm{D}_2$ transition of ${}^{113}\mathrm{{Cd}}^{+}$. The temperature obtained for ${}^{113}\mathrm{{Cd}}^{+}$ ions is 41(4) mK.}
\end{figure}

The temperature of SC ions is an important parameter in the sympathetic cooling scheme. We obtain the temperature of SC ${}^{113}\mathrm{{Cd}}^{+}$ ions by measuring the Doppler broadening of the laser cooling transition. The power of the 214-nm probe laser beam is maintained below $10\mu\mathrm{W}/\mathrm{mm}^2$, which avoids heating or cooling the ions. The graph of Fig .\ref{fig:fig_05} shows a typical measurement result. The frequency of the 214-nm probe laser was scanned around the resonant frequency over a range of 1.2 GHz. The measured line profile was fitted with a Voigt function. The Lorentz linewidth of the Voigt function was set to 60.13 MHz, which is the natural linewidth of the $\mathrm{D}_2$ transition of ${}^{113}\mathrm{{Cd}}^{+}$. The fitted Gaussian linewidth reflects the velocity distribution of ${}^{113}\mathrm{{Cd}}^{+}$ ions. Then the temperature is determined using \cite{drewsen2002ion,warrington2002temperature,riehle2006frequency}
\begin{equation}
\label{equ03}
T=\frac{M c^{2}}{8 \ln 2 \cdot k_{\mathrm{B}}}\left(\frac{v_{{G}}}{v_{0}}\right)^{2},
\end{equation}
where $c$ denotes the speed of light in vacuum, $k_{\mathrm{B}}$ the Boltzmann constant, $v_{{G}}$ the Gaussian linewidth, and $v_{0}$ the resonant frequency of the $\mathrm{D}_2$ transition of ${}^{113}\mathrm{{Cd}}^{+}$. In Fig. \ref{fig:fig_05}, the fitted Gaussian linewidth is 19.1(0.9) MHz. Therefore, the temperature of SC ${}^{113}\mathrm{{Cd}}^{+}$ ions is estimated to be 41(4) mK.

Moreover, we determine the loss rates of the two species of ions under the sympathetic cooling scheme by measuring the decay of the PMT counts respectively. In the few hours of measurement, the fluorescence counts of ${}^{40}\mathrm{{Ca}}^{+}$ ions (397 nm) did not decrease significantly, and the lifetime was estimated to be more than one day. This is easy to explain because the ${}^{40}\mathrm{{Ca}}^{+}$ ions are located in the center of the trap and are less affected by RF heating. As for ${}^{113}\mathrm{{Cd}}^{+}$ ions, the fluorescence counts (214 nm) followed exponential decay and the lifetime ($1/e$) was estimated to be about $2\times10^4$ seconds, which is three times longer than that in the laser cooling scheme\cite{miao2021precision}.


\section{Characterization of ion Coulomb crystals}

\begin{figure}[t!]
\centering
\includegraphics[width=8cm]{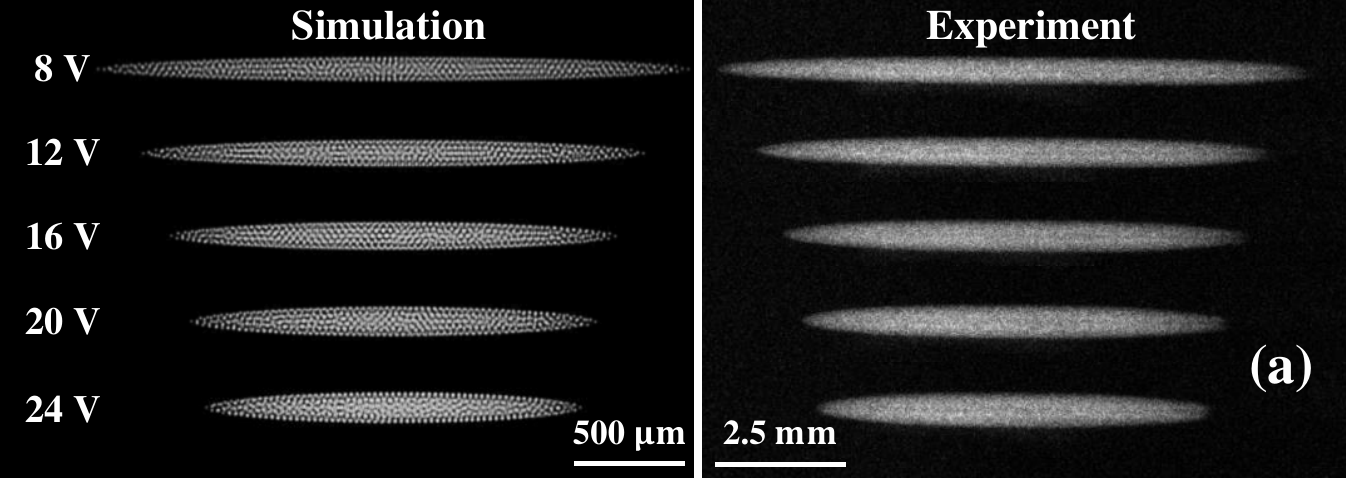} \\
\includegraphics[width=8cm]{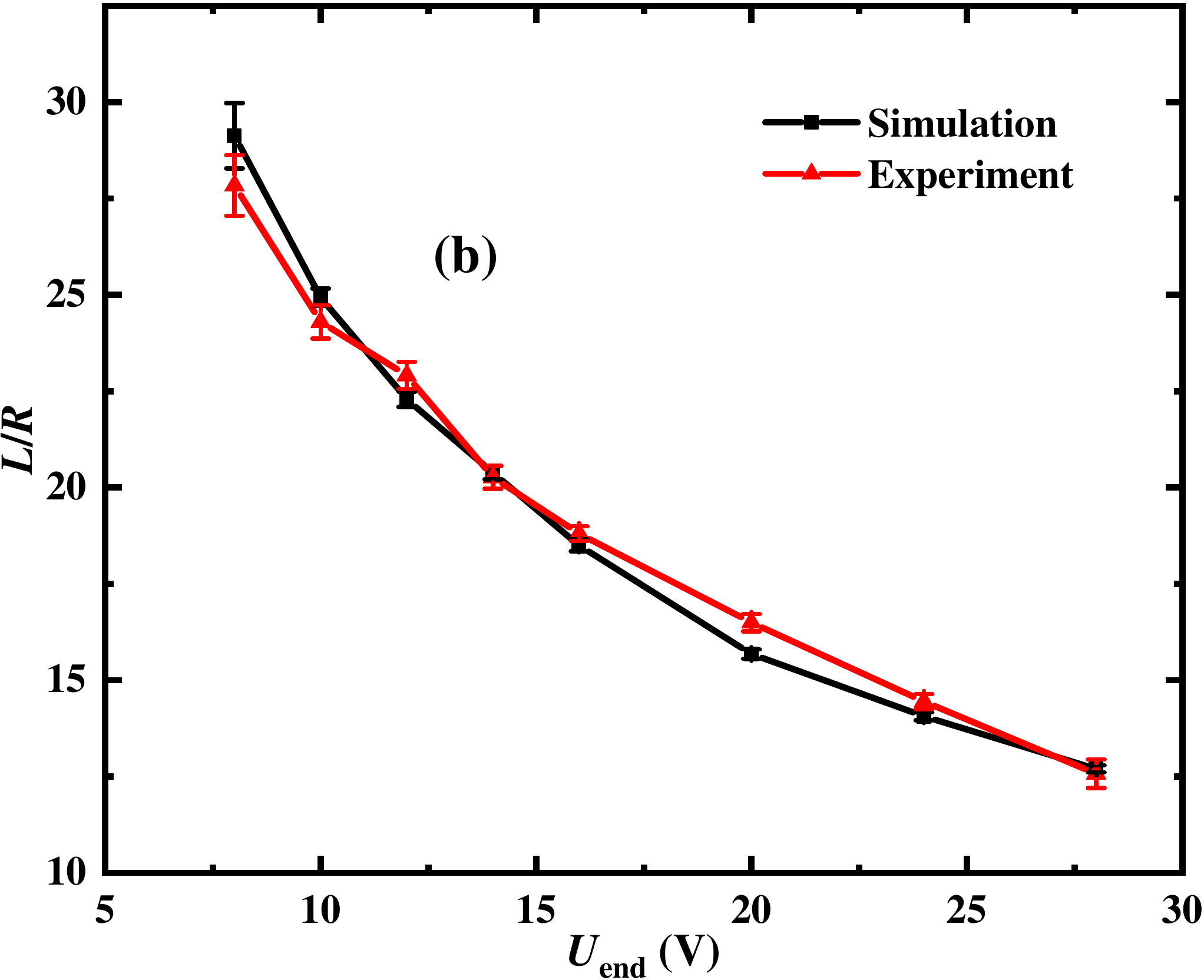}
\caption{\label{fig:fig_06} (Color online) (a) Simulated and experimental images of the ${}^{40}\mathrm{{Ca}}^{+}$ crystal for different DC voltages under the influence of a fixed RF voltage $U_{\mathrm{rf}}=300$ V, respectively. The number of ${}^{40}\mathrm{{Ca}}^{+}$ ions in simulations is 1024 but that in experiments is $6.5(0.2)\times10^4$. (b) The ion crystal aspect ratio $L/R$ (see text) of a spheroid fit to the ${}^{40}\mathrm{{Ca}}^{+}$ envelope for simulations and experiments under variable DC voltages.}
\end{figure}

With ultracold samples in the trap, we have studied the properties and the structures of the Coulomb crystals containing single species and double species of ions. In Fig. \ref{fig:fig_06}(a), we show simulated and experimental images of the ${}^{40}\mathrm{{Ca}}^{+}$ crystals for different DC voltages under the influence of a fixed RF voltage $U_{\mathrm{rf}}=300$ V, respectively. The number of ${}^{40}\mathrm{{Ca}}^{+}$ ions in simulations is 1024, whereas that in experiments is $6.5(0.2)\times10^4$. As can be seen, under the same electrical parameters, the shape of the ${}^{40}\mathrm{{Ca}}^{+}$ crystal obtained by simulating a small number of ions is similar to that formed by a large number of ions in the experiment. In Fig. \ref{fig:fig_06}(b), we present the crystal aspect ratio, i.e., the ratio between half length and radius, $L/R$, of a spheroid fit to the ${}^{40}\mathrm{{Ca}}^{+}$ envelope, for simulations and experiments under varying DC voltages. Comparing the experiment with the MD simulation, it is clear that the aspect ratio of the ion crystal is related to electrical parameters, but independent of the number of ions. Therefore, ion crystals with the same aspect ratios as those in the experiment can be obtained by performing MD simulations with a small number of ions. In addition, this can be used to determine the value of $\kappa$ in Eq. (\ref{equ02}) in the simulation model by comparing the simulation images with the experimental images, especially for low DC voltages \cite{okada2010characterization}.


\begin{figure}[t!]
\centering
\includegraphics[width=8cm]{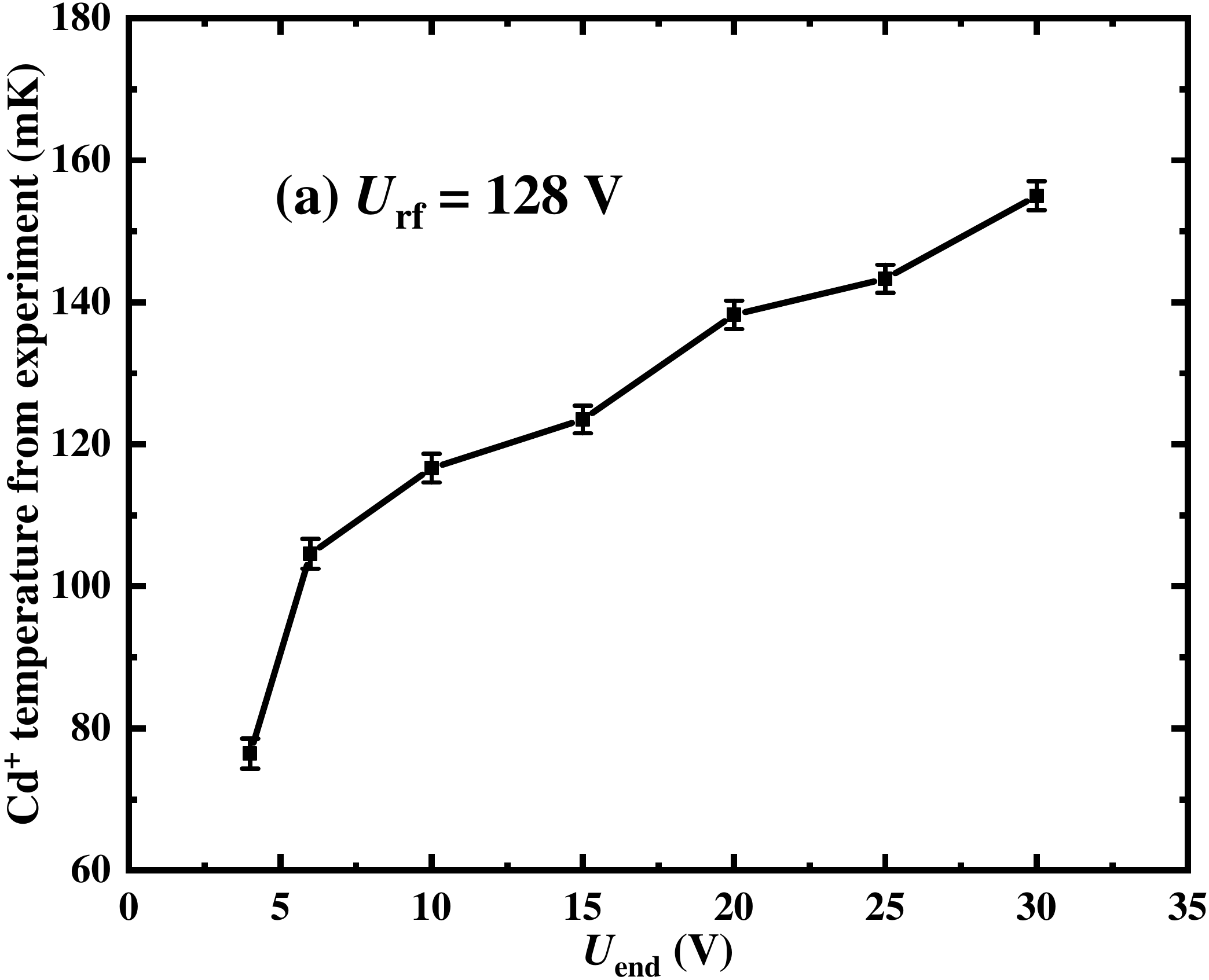} \\
\includegraphics[width=8cm]{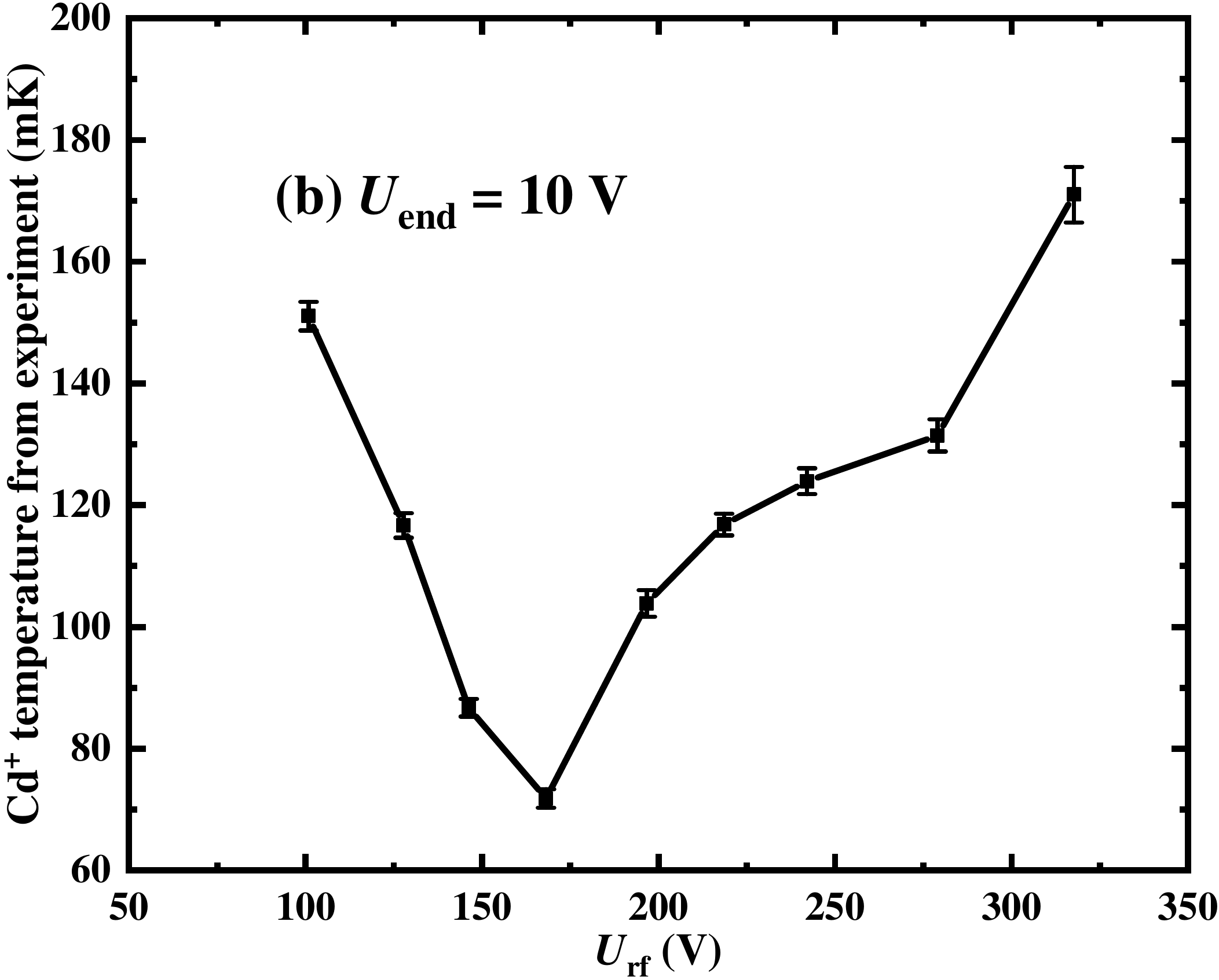}
\caption{\label{fig:fig_07} (Color online) Influence of electrical parameters on the temperature of SC ${}^{113}\mathrm{{Cd}}^{+}$ ions. (a) Temperatures at different DC voltages with $U_\mathrm{rf}=128$ V. (b) Temperatures at different RF voltages with $U_\mathrm{end}=10$ V. In our experiment, the optimal electrical parameters are $U_\mathrm{end}=4$ V and $U_\mathrm{rf}=170$ V, respectively.}
\end{figure}

The sympathetic cooling effect is the most important factor to be considered in order to achieve stable two-component Coulomb crystals. There are many factors that affect the sympathetic cooling effect such as electrical parameters, ion mass ratio and number ratio between LC and SC ions. In this paper, the sympathetic cooling effect is defined as the equilibrium temperature of SC ions, and the better effect leads to lower temperature.

First, the influence of electrical parameters on the temperature of SC ions is studied. In Fig. \ref{fig:fig_07}, we show the dependence of the temperature of outer SC ${}^{113}\mathrm{{Cd}}^{+}$ ions on variable electrical parameters. All the raw data in the graphs were obtained from the same ion bicrystal in a single loading. Each point is the average of 10 measurements to ensure the reliability of the data. 
The graph of Fig. \ref{fig:fig_07}(a) shows the relationship between the temperature of SC ${}^{113}\mathrm{{Cd}}^{+}$ ions and $U_{\mathrm{end}}$ under the influence of a fixed RF voltage $U_{\mathrm{rf}}=128$ V. It can be shown that the temperature of ${}^{113}\mathrm{{Cd}}^{+}$ ions and $U_{\mathrm{end}}$ are positively correlated. The same phenomenon was observed in the experiment \cite{xin2021research}. This is a result that the increase of $U_{\mathrm{end}}$ leads to an increase of the ion radial size, which intensifies the RF heating effect. It is worth noting that there is a lower limit for the DC voltage. In our experiment, $U_{\mathrm{end}}$ should be greater than 4 V in order to trap ions stably. Apart from that, the first-order Doppler frequency shift of trapped ions can be suppressed by reaching the Lamb-Dicke criterion, which requires the axial size of the ion crystal to be less than 50 mm ($2L\textless50$ mm). This also limits the minimum value of $U_{\mathrm{end}}$. However, for the RF voltage, there is an optimal value, corresponding to the lowest temperature of SC ${}^{113}\mathrm{{Cd}}^{+}$ ions, which is shown in Fig. \ref{fig:fig_07}(b). On the one hand, increasing $U_{\mathrm{rf}}$ compresses the radial size of the ion crystal, so as to reduce RF heating of trapped ions. On the other hand, the value of $q$ in Eq. (\ref{equ01}) increases with the increase of $U_{\mathrm{rf}}$, resulting in the increase of ion micromotion energy. These two effects have opposite effects on the ion temperature, which leads to an optimal value of $U_{\mathrm{rf}}$. In our experiment, the optimal $U_{\mathrm{rf}}$ is about 170 V.


\begin{figure}[t!]
\centering
\includegraphics[width=8cm]{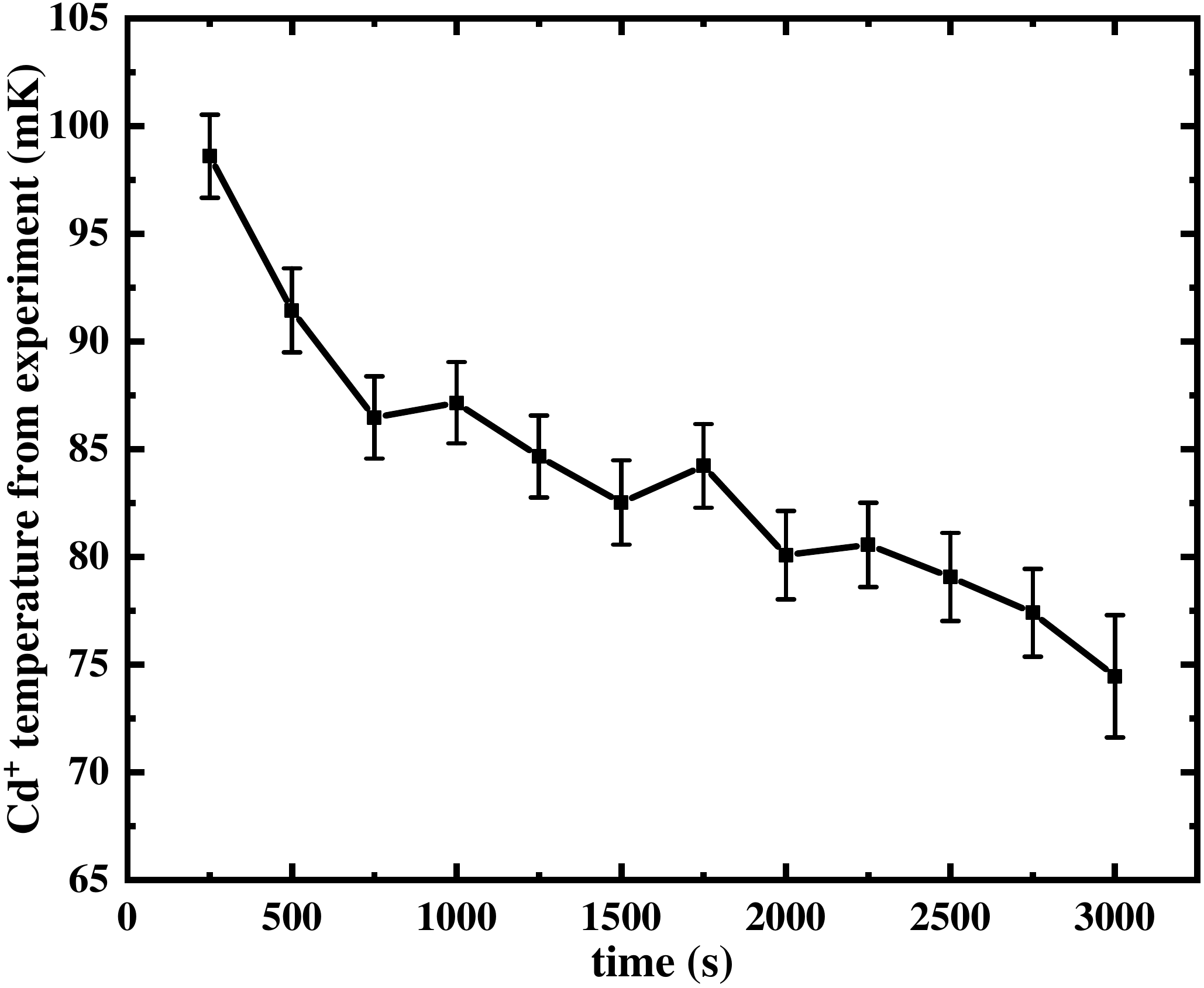}
\caption{\label{fig:fig_08} (Color online) Changing of the temperature of SC ${}^{113}\mathrm{{Cd}}^{+}$ ions with time under the sympathetic cooling scheme. In the experiment, the electrical parameters are fixed to $U_{\mathrm{rf}}=155$ V and $U_{\mathrm{end}}=5$ V. The temperature of ${}^{113}\mathrm{{Cd}}^{+}$ ions is obtained every 250 s. The uncertainty of each point comes from the fitting with a Voigt function.}
\end{figure}

\begin{figure}[t!]
\centering
\includegraphics[width=8cm]{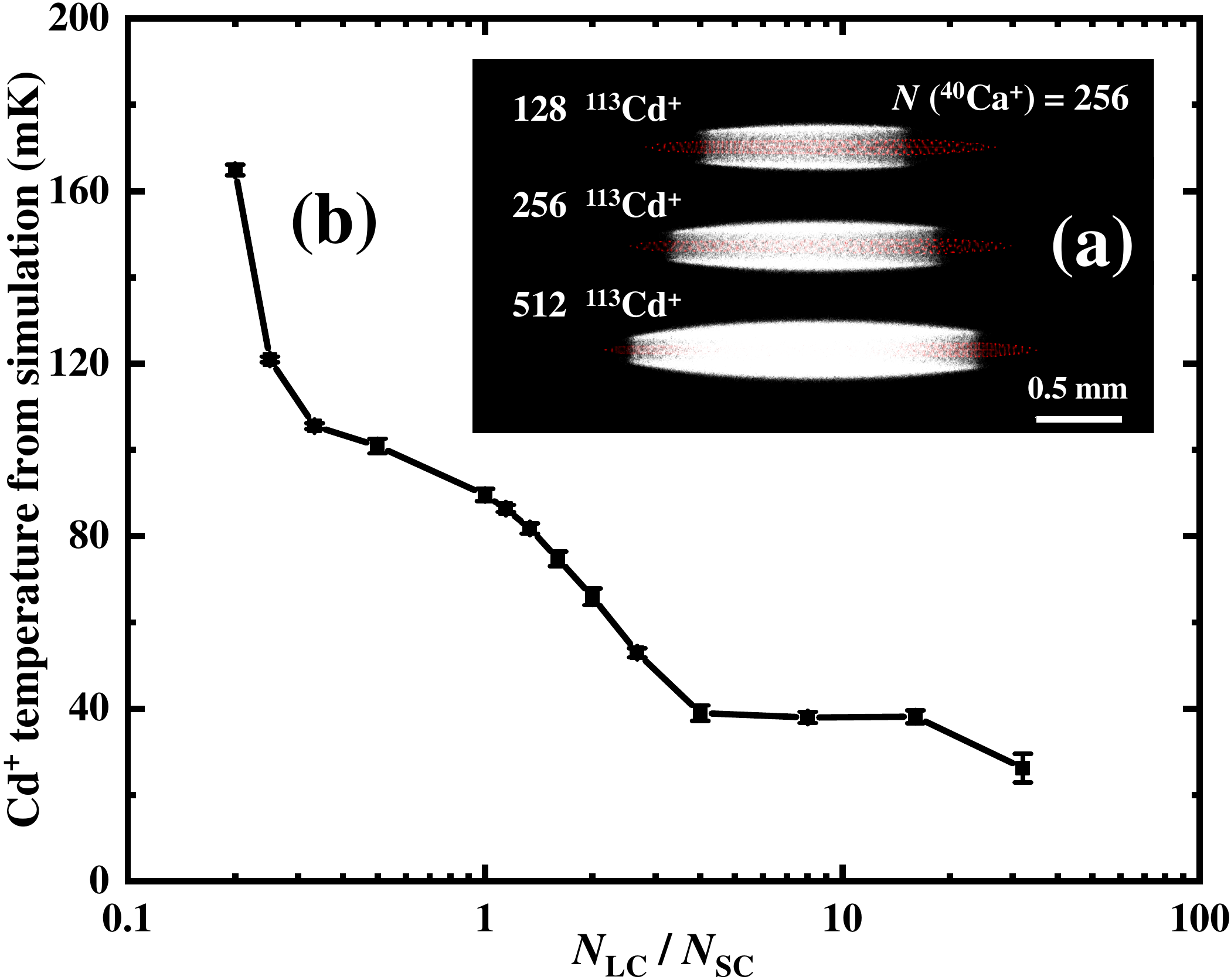}
\caption{\label{fig:fig_09} (Color online) (a) Simulation images for the sympathetic cooling for different numbers of ${}^{113}\mathrm{{Cd}}^{+}$ ions under the influence of a fixed number of ${}^{40}\mathrm{{Ca}}^{+}$ ions. During the simulation, the electrical parameters are the same as Fig. \ref{fig:fig_08}. The number of ${}^{40}\mathrm{{Ca}}^{+}$ ions is fixed to 256. (b) Relationship between the temperature of SC ${}^{113}\mathrm{{Cd}}^{+}$ ions and the number ratio of $N_{\mathrm{{Ca}}^{+}}/N_{\mathrm{{Cd}}^{+}}$.}
\end{figure}

The number ratio between LC and SC ions is also an important factor affecting the temperature of SC ions. However, there are few relevant studies because of the complexity to precisely control the number ratio in the experiment \cite{okada2010characterization}. For a large two-component ion crystal in our experiment, it is difficult to precisely control the number ratio between ${}^{40}\mathrm{{Ca}}^{+}$ and ${}^{113}\mathrm{{Cd}}^{+}$ ions. However, the loss rate of ${}^{113}\mathrm{{Cd}}^{+}$ ions from the ion trap is much higher than that of ${}^{40}\mathrm{{Ca}}^{+}$ ions because they are closer to the edge of the potential well under the sympathetic cooling scheme. Therefore, the relationship between the temperature of outer SC ${}^{113}\mathrm{{Cd}}^{+}$ ions and the number ratio of $N_{\mathrm{{Ca}}^{+}}/N_{\mathrm{{Cd}}^{+}}$ can be roughly described by continuously measuring the temperature of ${}^{113}\mathrm{{Cd}}^{+}$ ions for a long time. The results are shown in Fig. \ref{fig:fig_08}. During the measurement, the electrical parameters are fixed to $U_{\mathrm{rf}}=155$ V and $U_{\mathrm{end}}=5$ V. The temperature measurement process lasts about 1 hour. Compared with ${}^{113}\mathrm{{Cd}}^{+}$ ions, the decrease of ${}^{40}\mathrm{{Ca}}^{+}$ ions before and after the measurement can be ignored. The temperature of ${}^{113}\mathrm{{Cd}}^{+}$ ions is obtained every 250 s. The uncertainty of each point comes from the fitting with a Voigt function. The graph of Fig. \ref{fig:fig_08} shows that the temperature of ${}^{113}\mathrm{{Cd}}^{+}$ ions decreases with time, which indicates that the sympathetic cooling effect is better for small number of ${}^{113}\mathrm{{Cd}}^{+}$ ions when the number of ${}^{40}\mathrm{{Ca}}^{+}$ ions is fixed. This is also verified in MD simulation. In Fig. \ref{fig:fig_09}(a), we show simulated images for the sympathetic cooling for different numbers of ${}^{113}\mathrm{{Cd}}^{+}$ ions under the influence of a fixed number of ${}^{40}\mathrm{{Ca}}^{+}$ ions. The electrical parameters in the simulation are consistent with those in the experiment. The number of ${}^{40}\mathrm{{Ca}}^{+}$ ions is set to 256 to form a three-dimensional ellipsoidal structure. Figure \ref{fig:fig_09}(b) shows the temperature of SC ${}^{113}\mathrm{{Cd}}^{+}$ ions for different number ratios of $N_{\mathrm{{Ca}}^{+}}/N_{\mathrm{{Cd}}^{+}}$, which are consistent with the experimental results in Fig. \ref{fig:fig_08}.

\begin{figure}[t!]
\centering
\includegraphics[width=8cm]{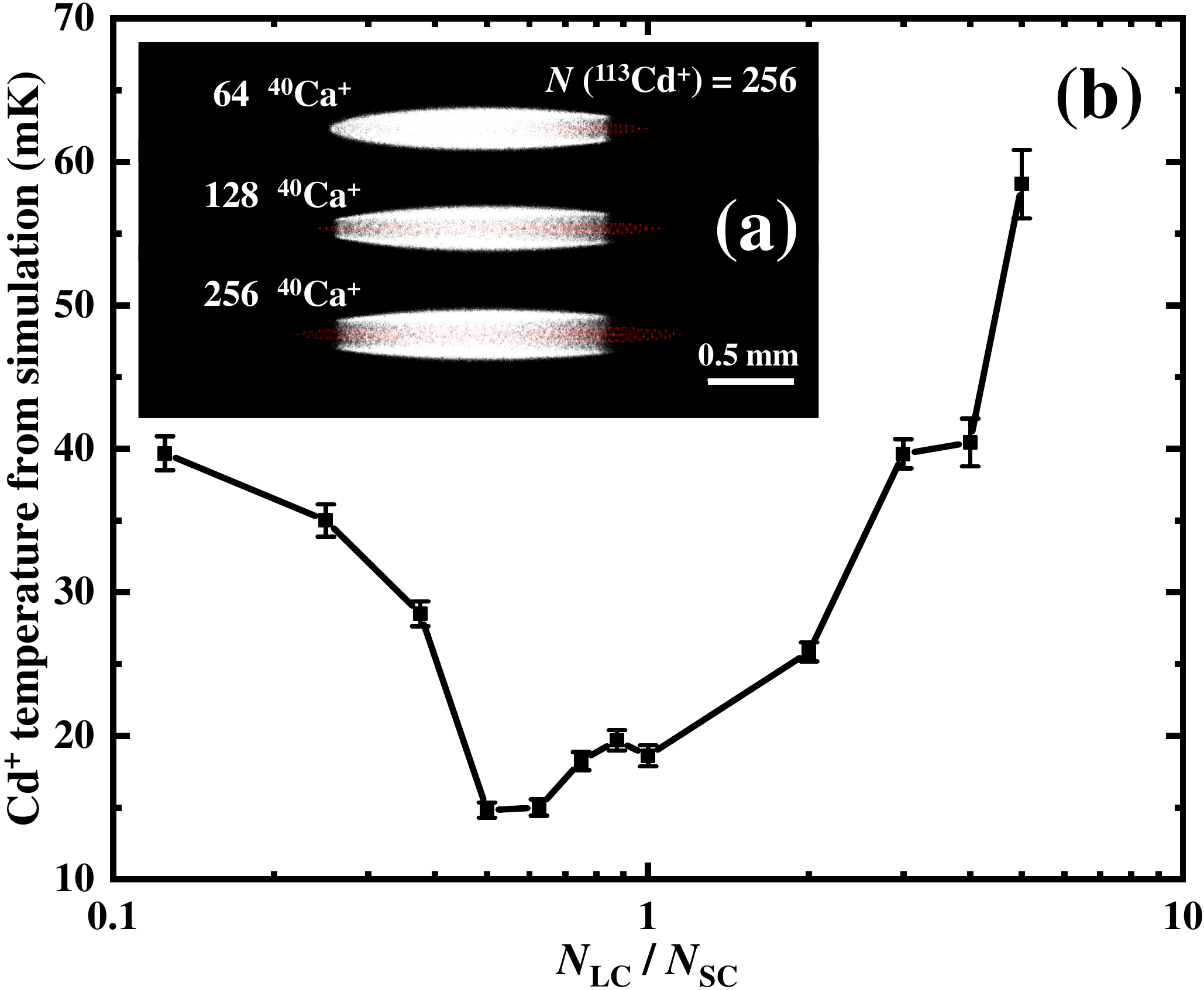}
\caption{\label{fig:fig_10} (Color online) (a) Simulation images for the sympathetic cooling for different numbers of ${}^{40}\mathrm{{Ca}}^{+}$ ions under the influence of a fixed number of ${}^{113}\mathrm{{Cd}}^{+}$ ions. During the simulation, the electrical parameters are the same as Fig. \ref{fig:fig_08}. The number of ${}^{113}\mathrm{{Cd}}^{+}$ ions is fixed to 256. (b) Relationship between the temperature of SC ${}^{113}\mathrm{{Cd}}^{+}$ ions and the number ratio of $N_{\mathrm{{Ca}}^{+}}/N_{\mathrm{{Cd}}^{+}}$.}
\end{figure}

On the other hand, there is an optimal value of $N_{\mathrm{{Ca}}^{+}}/N_{\mathrm{{Cd}}^{+}}$ to maximize the sympathetic cooling effect for varying $N_{\mathrm{{Ca}}^{+}}$ and $N_{\mathrm{{Cd}}^{+}}=256$, as shown in Fig. \ref{fig:fig_10}(b). As the number of ${}^{40}\mathrm{{Ca}}^{+}$ ions increases, the ${}^{113}\mathrm{{Cd}}^{+}$ ions are repulsed to the edge of the potential well (see Fig. \ref{fig:fig_10}(a)). The optimal number ratio between ${}^{40}\mathrm{{Ca}}^{+}$ and ${}^{113}\mathrm{{Cd}}^{+}$ ions is approximately 0.5, which differs from the results in Fig. \ref{fig:fig_09} even though the ratio of $N_{\mathrm{{Ca}}^{+}}/N_{\mathrm{{Cd}}^{+}}$ is the same. In summary, the sympathetic cooling effect is therefore dependent on the total number of ions ($N_{\mathrm{{Ca}}^{+}}+N_{\mathrm{{Cd}}^{+}}$) as well as the ratio.

\section{Conclusion}
In conclusion, we have sympathetically cooled and crystallized ${}^{113}\mathrm{{Cd}}^{+}$ ions using laser-cooled ${}^{40}\mathrm{{Ca}}^{+}$ ions, and directly observed the complete large bicrystal structure in a linear Paul trap for the first time. The large Coulomb crystal contains up to approximately $3.5\times10^3$ ${}^{40}\mathrm{{Ca}}^{+}$ ions and $6.8\times10^3$ ${}^{113}\mathrm{{Cd}}^{+}$ ions. Among them, the lighter ${}^{40}\mathrm{{Ca}}^{+}$ ions form an ellipsoidal shell structure along the axis. However, the envelope of the heavier ${}^{113}\mathrm{{Cd}}^{+}$ ions is related to the number ratio between the two types of ions. Assuming that the number of ${}^{40}\mathrm{{Ca}}^{+}$ ions is constant, the ${}^{113}\mathrm{{Cd}}^{+}$ ions form a thin shell outside of the ${}^{40}\mathrm{{Ca}}^{+}$ crystal when the number of ${}^{113}\mathrm{{Cd}}^{+}$ ions is small. As the number of ${}^{113}\mathrm{{Cd}}^{+}$ ions increases, the structure of the ${}^{113}\mathrm{{Cd}}^{+}$ crystal transforms into a hollow ellipsoid. Under the sympathetic cooling scheme, the temperature of the crystallized ${}^{113}\mathrm{{Cd}}^{+}$ ions was measured to be as low as 41 mK with a large mass ratio.

With ultracold ion samples in the trap, we have studied several properties and structures of the Coulomb crystals containing single species and double species of ions. Comparing the experiment with the MD simulation, it is clear that the aspect ratio of the ion crystal is related to electrical parameters, whereas independent of the number of ions. Therefore, ion crystals with the same aspect ratios as those in the experiment can be obtained by performing MD simulations with a small number of ions. 
In order to maximize the sympathetic cooling efficiency limited by the large mass ratio of $M_{\mathrm{{Cd}}^{+}}/M_{\mathrm{{Ca}}^{+}}=2.825$, the factors affecting the sympathetic cooling effect have been studied in detail, including the electrical parameters and the number ratio between laser-cooled ${}^{40}\mathrm{{Ca}}^{+}$ ions and sympathetically-cooled ${}^{113}\mathrm{{Cd}}^{+}$ ions. In our experiment, the optimal electrical parameters are determined to be $U_\mathrm{end}=4$ V and $U_\mathrm{rf}=170$ V, respectively. 
For the optimal number ratio between the two types of ions, both experiments and simulations are performed, yielding a conclusion that the sympathetic cooling effect is dependent on the total number of ions as well as the ratio. When the number of ${}^{40}\mathrm{{Ca}}^{+}$ ions is fixed, there is a positive correlation between the temperature and the number of ${}^{113}\mathrm{{Cd}}^{+}$ ions. When the number of ${}^{113}\mathrm{{Cd}}^{+}$ ions is fixed, there is an optimal number ratio of about 0.5 between laser-cooled ${}^{40}\mathrm{{Ca}}^{+}$ ions and sympathetically-cooled ${}^{113}\mathrm{{Cd}}^{+}$ ions, corresponding to the lowest temperature of the ${}^{113}\mathrm{{Cd}}^{+}$ crystal.

It is worth mentioning that the experiment of sympathetic cooling ${}^{40}\mathrm{{Ca}}^{+}$ ions with laser-cooled ${}^{113}\mathrm{{Cd}}^{+}$ ions has also been done in our laboratory. The mass ratio between sympathetically-cooled and laser-cooled ions is 0.354, which is extremely challenging, especially for the mass ratio lower than 0.52 \cite{baba2002sympathetic,schiller2003molecular}. However, the temperature of the crystallized ${}^{40}\mathrm{{Ca}}^{+}$ ions was measured to be approximately 130 mK, which is higher than that of ${}^{113}\mathrm{{Cd}}^{+}$ sympathetically-cooled by ${}^{40}\mathrm{{Ca}}^{+}$. This may be explained by the fact that the small mass ratio of $M_{\mathrm{SC}}/M_{\mathrm{LC}}$ limits the sympathetic cooling efficiency. 

The ${}^{40}\mathrm{{Ca}}^{+}$-${}^{113}\mathrm{{Cd}}^{+}$ cooling and crystallization results in this paper enrich the relevant experimental research of large two-component ion crystals, especially for the properties and structures of the large Coulomb crystals. Moreover, the ultracold sample of ${}^{113}\mathrm{{Cd}}^{+}$ makes it possible to further improve the accuracy of the cadmium-ion microwave clock, which may prove to be useful as a highly accurate, portable atomic clock.

\begin{acknowledgments}
The authors thank C. F. Wu, J. Z. Han, H. X. Hu, W. X. Shi, T. G. Zhao and Y. Zheng for helpful assistance and discussions. This work is supported by National Natural Science Foundation of China (12073015), Beijing Natural Science Foundation (1202011), and Tsinghua University Initiative Scientific Research Program.
\end{acknowledgments}

\appendix


\bibliography{apssamp}

\end{document}